\documentclass[namedreferences,hyperref,optionalrh]{spr-sola}
\usepackage{graphicx}        % For eps figures, newer & more powerful
\usepackage{amssymb}        % useful mathematical symbols
\usepackage{color}           % For color text: \color command
\usepackage{siunitx} %forSIunit
\usepackage{booktabs}
\usepackage{multirow}
\usepackage{graphicx}
\usepackage{amsmath}
\usepackage{tabularx}
\usepackage{enumitem}
\usepackage[rightcaption]{sidecap}
\chardef\us=`\_

\begin{document}

\begin{frontmatter}
\title{Investigating the Relationship Between Physical Properties and Spatial Irregularities at Coronal Hole Boundaries}

\author[addressref={aff1,aff2}, corref,email={ngampoopun@mps.mpg.de}]{\inits{N.}\fnm{Nawin}~\snm{Ngampoopun}\orcid{0000-0002-1794-1427}}
\author[addressref={aff3,aff4}, email={david.long@dcu.ie}]{\inits{D.M.}\fnm{David~M.}~\snm{Long}\orcid{0000-0003-3137-0277}}
\author[addressref=aff2, email={lucie.green@ucl.ac.uk}]{\inits{L.M.}\fnm{Lucie~M.}~\snm{Green}\orcid{0000-0002-0053-4876}}
\author[addressref={aff5,aff2,aff6}, email={steph.yardley@northumbria.ac.uk}]{\inits{S.L.}\fnm{Stephanie~L.}~\snm{Yardley}\orcid{0000-0003-2802-4381}}
\author[addressref=aff2, email={alexander.james@ucl.ac.uk}]{\inits{A.W.}\fnm{Alexander~W.}~\snm{James}\orcid{0000-0001-7927-9291}}
\author[addressref=aff7, email={emason@predsci.com}]{\inits{E.I.}\fnm{Emily~I.}~\snm{Mason}\orcid{0000-0002-8767-7182}}
\author[addressref={aff8,aff9}, email={stephan.heinemann@helsinki.fi}]{\inits{S.G.}\fnm{Stephan~G.}~\snm{Heinemann}\orcid{0000-0002-2655-2108}}
\author[addressref={aff10,aff11}, email={uritsky@cua.edu}]{\inits{V.M.}\fnm{Vadim~M.}~\snm{Uritsky}\orcid{0000-0002-5871-6605}}

%\author{\inits{}\fnm{}~\lnm{}\orcid{}}
%   NOTE:  Just one corresponding author [corref]
\address[id=aff1]{Max Planck Institute for Solar System Research, Justus-von-Liebig-Weg 3, 37077 G{\"o}ttingen, Germany}
\address[id=aff2]{University College London, Mullard Space Science Laboratory, Holmbury St. Mary, Dorking, Surrey, RH5 6NT, UK}
\address[id=aff3]{Centre for Astrophysics \& Relativity, School of Physical Sciences, Dublin City University, Glasnevin Campus, Dublin, D09 V209, Ireland}
\address[id=aff4]{Astronomy \& Astrophysics Section, Dublin Institute for Advanced Studies, Dublin D02 XF86, Ireland}
\address[id=aff5]{Department of Mathematics, Physics and Electrical Engineering, Northumbria University, Ellison Place, Newcastle Upon Tyne, NE1 8ST, UK}
\address[id=aff6]{Donostia International Physics Center (DIPC), Paseo Manuel de Lardizabal 4, 20018, San Sebastián, Spain}
\address[id=aff7]{Predictive Science Inc., San Diego, CA 92121, USA}
\address[id=aff8]{Department of Physics, University of Helsinki, 00014, Helsinki, Finland}
\address[id=aff9]{Institute of Physics, University of Graz, Universit{\"a}tsplatz 3, Graz, Austria}
\address[id=aff10]{Catholic University of America, 620 Michigan Avenue, N.E. Washington, DC 20064, USA}
\address[id=aff11]{NASA Goddard Space Flight Center, 8800 Greenbelt Avenue, Greenbelt, MD 20771, USA}

\runningauthor{Ngampoopun et al.}
\runningtitle{Properties of CH boundary}

\begin{abstract}
% The origin of the solar wind remains an open question in solar and heliospheric physics. One proposed scenario is that some slow solar wind may be released at coronal hole boundaries via magnetic reconnection, a process that also dominates the evolution of coronal hole boundaries. 
 % {\bf --- Modified slightly from ApJ version}
Coronal hole boundaries are the interfaces between closed and open magnetic field regions in the solar atmosphere. Many fundamental processes take place at these regions, including magnetic reconnection that is responsible for solar wind release and restructuring of the solar magnetic field.
In this paper, we present a case study in which we investigate the physical properties of the boundary of a large low-latitude coronal hole. Differential Emission Measure analysis is used to derive the plasma properties of these regions. We also apply correlation dimension mapping analysis to measure the irregularities of the coronal hole boundary. We find that the leading boundary of this coronal hole has a slightly higher average plasma temperature, is associated with a stronger and more unipolar magnetic field, and has a smoother boundary line than the trailing counterpart. These differences are hypothesised to be direct consequences of the local magnetic field configurations of the coronal hole boundary: the leading boundary corresponds to large, well-organised coronal loops, and the trailing boundary corresponds to more dispersed, randomly orientated small magnetic bipoles. Hence, we suggest that the surrounding magnetic field structure and the nature of magnetic reconnection influence the properties of coronal hole boundaries.
%The differences in the magnetic field structure in those regions may also influence the nature of interchange reconnection.
\end{abstract}
\end{frontmatter}
%-------------------------------------------------

\section{Introduction}
     \label{S-Introduction}

Coronal holes (CHs) are long-lived regions of reduced extreme ultraviolet (EUV) and X-ray emission in the solar corona. They are characterised by relatively low plasma temperature (0.8 -- 1 MK), low plasma density ($1-3 \times10^{-8}~\text{cm}^{-3}$), and weak magnetic field strength (1 -- 5 G) compared to the surrounding quiet Sun \citep[see e.g.,][]{Cranmer2009, Heinemann2019, Heinemann2021}. While CHs are primarily located in the polar regions during solar minimum, they are often found at lower latitudes at other phases of solar cycles. The formation of these low-latitude CHs is thought to be related to active regions (ARs) \citep{Wang2010, Karachik2010, Petrie2013} or eruptions of solar filaments \citep{Heinemann2018b, Hofmeister2025b}.

Due to their open magnetic field configuration, they are widely regarded as major sources of the fast solar wind. Meanwhile, the boundaries of CH have also been proposed as one of the sources of the slow solar wind \citep[see reviews by][]{Abbo2016, Viall2020}. At these boundaries, the slow solar wind plasma may travel along open magnetic fields with high expansion factors \citep[e.g.,][]{Wang1990, Wang2019} and/or it may be released through the interchange reconnection between the closed and open magnetic fields at various altitudes \citep[e.g.,][]{Fisk2003, Antiochos2011, Ngampoopun2025}. Interchange reconnection also plays a major role in the short-timescale evolution of CHs and their boundaries \citep{Baker2007, Kong2018, Heinemann2023}, as well as in maintaining the quasi-rigid rotation of CHs in the corona, in contrast to the photosphere below \citep{Timothy1975, Wang2004}. As this process takes place primarily in the corona, the evolution of CH areas may not necessarily correspond to the photospheric magnetic field beneath them \citep{Heinemann2020}.

Remote sensing observations have shown that bright points and jets are present within CHs and their boundary regions and are believed to be related to small-scale interchange reconnection and the evolution of CH boundaries \citep[e.g.,][]{Madjarska2009, Subramanian2010, Yang2011, Heinemann2023}. However, some changes in CH boundaries are not accompanied by obvious signatures in EUV or X-ray observations \citep[e.g.,][]{Kahler2002, Kahler2010}, maybe because reconnection produces insufficient energy to noticeably heat the plasma or reconnection occurs at much higher altitudes near the top of streamers \citep{Wang2000, Liewer2023}. 

Understanding the properties and dynamics of the CH boundary regions is of great interest, as they have significant implications on the resulting solar wind emanating from them and the structure of the heliosphere. The position, area and morphology of CHs show correlation with the peak solar wind speed \citep{Rotter2012, Samara2022}. In addition, the CH boundary regions that do not appear dark in EUV observations may contribute to the open flux problem \citep{Linker2017, Linker2021, Heinemann2024}, which is the observed mismatch between the amount of open magnetic flux derived from remote sensing and in situ observations by a factor of two or more. 

\citet{Heinemann2019} demonstrated that the intensity gradient across the boundary of low-latitude CHs is dependent on the surrounding magnetic field configuration (e.g. ARs, filaments, coronal loops) and as such on the solar cycle. This dependence consequently affects the optimal threshold and uncertainties for the extraction of the CH boundaries \citep{Reiss2021, Reiss2024}. 
Using Differential Emission Measure (DEM) analysis, \citet{Saqri2020} showed that there are clear differences in the DEM distribution between regions inside and outside the CH. Statistical analysis by \citet{Heinemann2021} further revealed that plasma density and emission measure change drastically at regions within $\sim15$ Mm (20\SI{}{\arcsecond}) of the CH boundaries, whereas plasma temperature variations in the same regions are much smaller. \citet{Koukras2025} investigated the plasma composition in the CH boundary regions by calculating the abundance ratio of the elements with low and high first ionisation potential (FIP). They found that these abundance ratios (i.e. FIP bias) increase monotonically from the CH boundaries until they reach the quiet Sun level at $\sim30-60$ Mm from the CH boundary.

In many cases, the CH boundaries appear to be spatially complex with a fuzzy or ragged appearance \citep{Kahler2002, Heinemann2020}. \citet{Aslanyan2022} proposed that the global magnetic field topology and dynamics directly influence the complexity of the CH boundaries, as their magnetohydrodynamic (MHD) simulation showed that the CH-helmet streamer boundary is more complex than the CH-pseudostreamer boundary. Their finding is further supported by the analysis presented by \citet{Mason2022}, who quantified the irregularity of the CH boundary lines observed in EUV images using the correlation dimension mapping (CDM) method. The authors also proposed  that the most relevant physical scale of the CH boundary is on the order of 5 -- 20 Mm, corresponding to small-scale plumelets and supergranulation. Lastly, \citet{Samara2022} found that the correlation between CH area and solar wind speed decreases significantly for CH with more complex boundaries.

However, it remains unclear how these spatial irregularities may be related to the observed physical properties (e.g., plasma temperature, magnetic flux) of the CH boundary regions. In this paper, we study the possible connection between plasma properties, magnetic field properties, and the spatial irregularity of the boundary of a long-lived low-latitude CH using high-cadence remote sensing observations from the Solar Dynamics Observatory \citep[SDO;][]{Pesnell2012}.
 % Differential emission measure (DEM) analysis is used to derive plasma properties at the CH boundary. CDM analysis is implemented to quantify the irregularities of the CH boundary line. Lastly, the magnetic field properties are derived from line-of-sight photospheric magnetograms.
The paper is outlined as follows. The datasets used in this study and the CH boundary extraction methods are described in Section~\ref{S-obs}. The structure of the extrapolated CH magnetic field is detailed in Section~\ref{S-mod}. In Section~\ref{S-res}, we present the results of the analysis of the properties of the CH boundary using DEM analysis on EUV images, the CDM method on the CH boundary line, and photospheric magnetograms. Finally, a discussion of the results and their implications is presented in Section~\ref{S-disc}.

\section{Remote Sensing Observations} \label{S-obs}
\begin{figure}[t]
    \centering
    \includegraphics[width = \columnwidth]{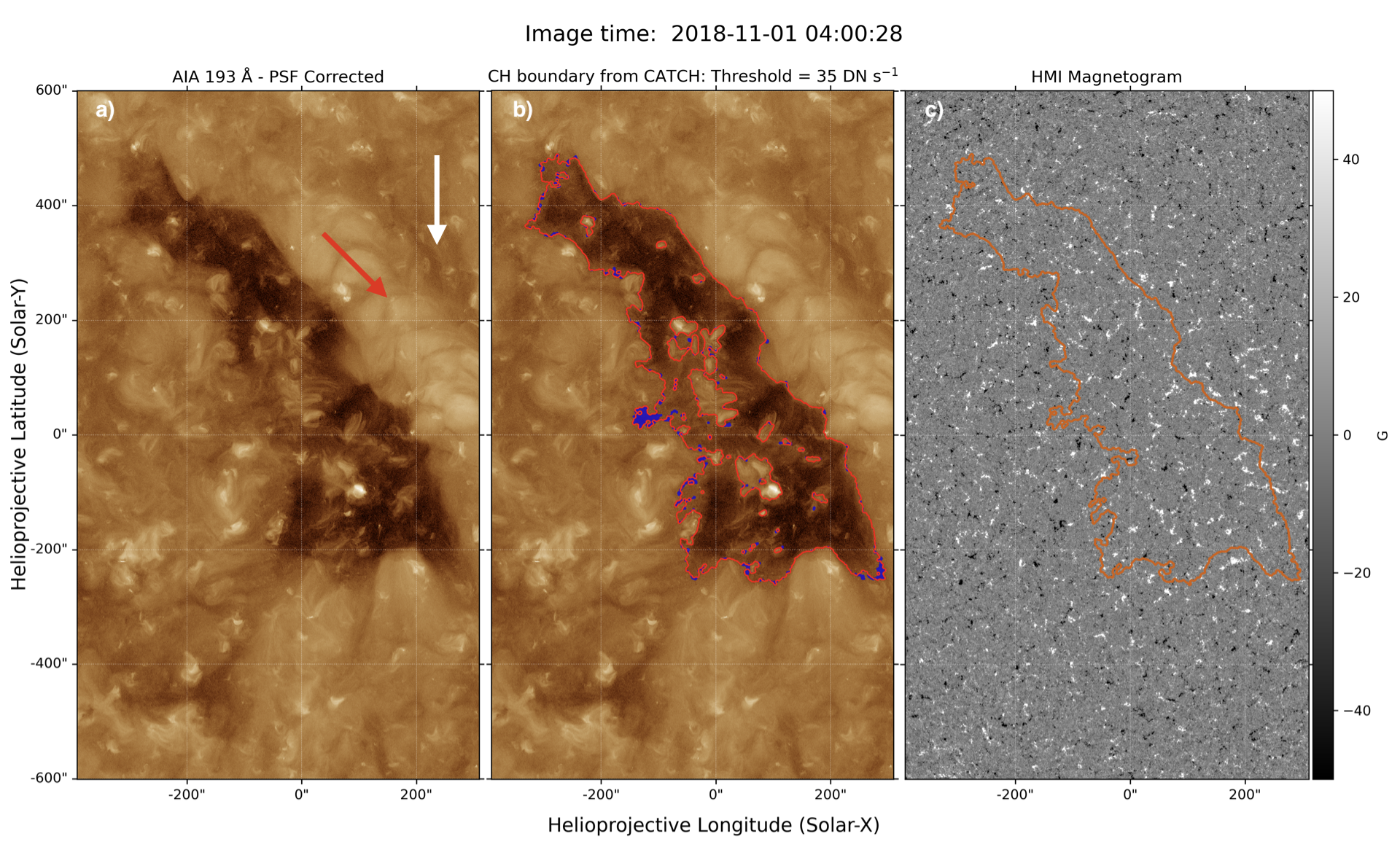}
    \caption{SDO observations of a low-latitude coronal hole on 2018 November 1 at 04:00:28 UT. Panel a displays the processed and PSF-corrected AIA 193~\AA\ (see Section~\ref{S-obs}).The white and red arrows indicate the filament channel and coronal cell structure, respectively. Panel b shows the CH boundary derived using the CATCH algorithm in a red contour, with the uncertainties of the boundary shown in blue. Panel c corresponds to HMI LOS photospheric magnetogram, with the final CH boundary contour overplotted in orange. The final CH boundary is obtained from filling the holes in the CATCH CH boundary contour.}
    \label{fig:AIA_HMI}
\end{figure}

On 2018 November 1, a large low-latitude coronal hole was observed on the solar disc from Earth.
%It is thought to be an extension of the northern polar CH and lasted approximately nine months \citep{Karna2022}. 
There were no emerging ARs or solar eruptions that occurred on that date that could significantly deform the CH boundary. Hence, this CH is a suitable target for this analysis, as we focused on the intrinsic properties of the CH boundary and its surrounding quiet Sun. To minimise the uncertainties of CH boundary detection due to line-of-sight (LOS) integration effect, we focus the analysis on a 1-hour period from November 1, 04:00 UT to 05:00 UT, coinciding with the time this CH passed the central meridian as observed by SDO.

EUV images in seven passbands, taken by the Atmospheric Imaging Assembly \citep[AIA;][]{Lemen2012} onboard SDO, were used to identify the CH and investigate its evolution. The plate scale of the instrument is 0.6\SI{}{\arcsecond}\ per pixel (equivalent to $\sim$~435 km) and the temporal cadence is 12~s. To mitigate stray and scattered light in the CH observations \citep{Saqri2020, Heinemann2021, Young2022}, the AIA images were deconvolved using the revised point spread function (PSF) for AIA derived by \citet{Hofmeister2025}. The deconvolution is performed using the Basic Iterative Deconvolution algorithm by \citet{Hofmeister2024}. The images were also registered, normalised to a one-second exposure time, and corrected for instrument degradation using the aiapy \citep{Barnes2020} Python library. These steps were performed on all AIA images prior to analysis.

The magnetic properties of the CH boundary were derived from LOS photospheric magnetograms obtained from the Helioseismic and Magnetic Imager \citep[HMI;][]{Scherrer2012}, also onboard SDO. The noise level is $\approx$ 7~G \citep{Couvidat2016} and the temporal cadence is 45~s. As the HMI plate scale is 0.505\SI{}{\arcsecond}\ per pixel, the HMI magnetograms were rescaled and coaligned with AIA images to have the same spatial resolution.

Figure \ref{fig:AIA_HMI} shows the cutouts of AIA 193~\AA\ (panels a and b) and HMI (panel c) observations of the low-latitude CH. There is a dark filament channel (shown by the white arrow in Figure \ref{fig:AIA_HMI}) present to the west of the CH, and there are several small coronal bright points located inside the CH and along its boundary. The dominant magnetic field in the CH is of positive (outward) polarity, as indicated by the LOS magnetogram in the right panel. We derived the CH boundary from the prepped AIA 193~\AA\ images using the Collection of Analysis Tools for Coronal Holes \citep[CATCH/pyCATCH;][]{Heinemann2019} algorithm. This method defined the optimal CH boundary threshold as the value at which the uncertainties of the CH area, obtained from varying the thresholds by small value ranges, are minimum. Hence, the derived boundary should correspond to locations between CH and the quiet Sun with maximum intensity gradients. For our case, we found that the threshold of 35 DN~s$^{-1}$ is optimal. The CH boundary and associated uncertainties obtained from the optimal value are shown in panel b of Figure~\ref{fig:AIA_HMI}. The uncertainties obtained from varying the threshold by $\pm$ 2 DN~s$^{-1}$. As we focused only on the leading and trailing boundaries of the CH, interior holes in the CH boundary contour caused by bright points inside the CHs were then filled using binary operators. The final CH boundary that we used in the analysis is then shown in panel c of Figure~\ref{fig:AIA_HMI}.

% It should be noted that small changes in quiet Sun intensity during solar minimum may also lead to considerable changes in the derived CH boundary. 
The derived CH boundary from CATCH should not be highly sensitive to small changes in quiet-Sun intensity. Still, to investigate the effect of varying the CH boundary threshold, we also derived the CH boundaries using the thresholds of 30 DN~s$^{-1}$ and 40 DN~s$^{-1}$ and discussed their implications in Appendix~\ref{S-A1}. In short, the small changes ($\pm$~5 DN~s$^{-1}$) in CH boundary extraction thresholds do not qualitatively affect our main conclusion. However, it should be noted that the derived CH properties can still vary significantly when using different CH detection schemes that have different underlying principles \citep[e.g.][]{Reiss2021, Reiss2024}.

% An intensity thresholding technique was used to extract the CH boundary from the EUV images shown in Figure \ref{fig:AIA_HMI}. The threshold value was defined as 46.2\% of the solar disc median intensity in the 193 \AA\ passband, corresponding to the mean threshold value for CH extraction using Collection of Analysis Tools for Coronal Holes \citep[CATCH;][]{Heinemann2019} during 2017--2019. This value was then applied to the 8-point median smoothed AIA 193 \AA\ cutout images to create a binary map at each time step, from which the CH boundary can be derived. 

\section{Global Magnetic Field Configuration} \label{S-mod}
\begin{figure*}[t!]
    \centering
    \includegraphics[width = 0.9\textwidth]{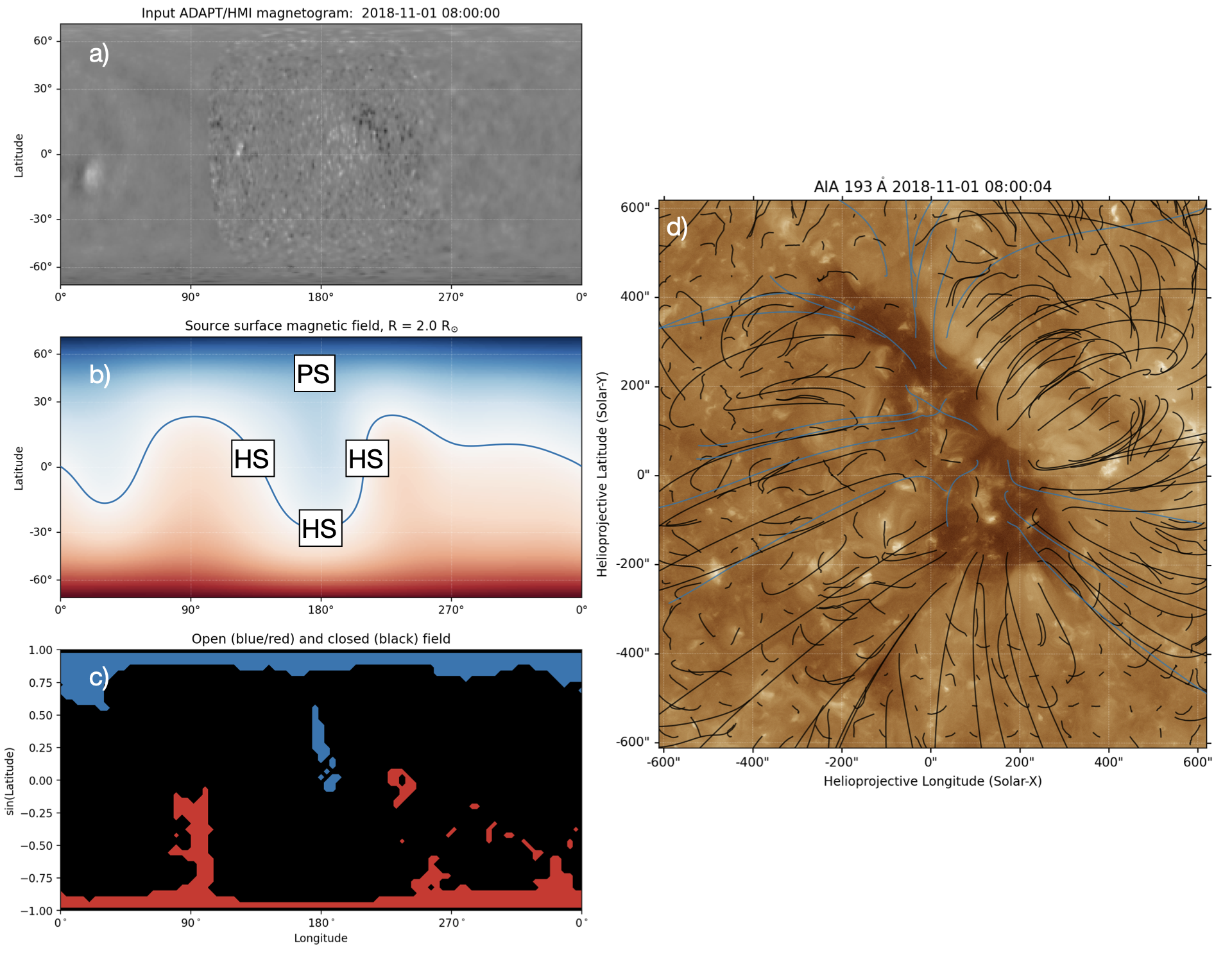}
    \caption{Global magnetic field configuration from the PFSS extrapolation. (a) The ADAPT/HMI magnetogram used as input for the PFSS extrapolation, saturated at $\pm50$~G. (b) The magnetic field polarity at the source surface of 2 R$_\odot$, saturated at $\pm0.5$~G. The blue line represents the heliospheric current sheet, and the location of helmet streamers (HS) and pseudostreamers (PS) are labelled. (c) The footpoints of positive (blue) and negative (red) open magnetic field lines. The low-latitude CH is located at 180$^\circ$ Carrington longitude. (d) The extrapolated magnetic field lines plotted over the zoomed-in AIA 193~{\AA} image of CH. The positive open field lines are plotted in blue, while the closed field lines are plotted in black.}
    \label{fig:PFSS_summary}
\end{figure*}

Potential field source surface (PFSS) extrapolations were performed to infer the magnetic configuration of the CH and the surrounding region. The Air Force Data Assimilative Photospheric Flux Transport \citep[ADAPT;][]{Arge2010, Hickmann2015} magnetograms, which incorporate a surface flux transport model to the HMI LOS magnetogram, were used for the boundary condition of the PFSS extrapolation. The transport model generated 12 realisations of the synoptic magnetogram, and the average of all realisations was used in this analysis. The pfsspy Python package\footnote{Now available as the sunkit-magex package:\url{https://docs.sunpy.org/projects/sunkit-magex/en/latest/}} \citep{Stansby2020b} was used to compute the PFSS solutions in a 3D grid equally spaced in sin(latitude), longitude, and heliocentric distance. The radius of the source surface (R$_{ss}$) was chosen to be 2.0 R$_\odot$. By comparing the magnetic field values from modelling and in-situ observations,  \citet{Badman2020} found that this R$_{ss}$ = 2.0 R$_\odot$ value is optimal for modelling the magnetic connectivity between the Parker Solar Probe \citep[PSP;][]{Fox2016} and the Sun during the first encounter, when the spacecraft was magnetically connected to this CH.

Figure \ref{fig:PFSS_summary} shows the summary of the PFSS extrapolation. The input ADAPT{/}HMI magnetogram is displayed in panel a. The higher resolution portion of the input map is the measurement from the HMI obtained near the CH passage over the central meridian, while the lower resolution portion corresponds to the magnetic field derived from the flux transport model. Note that the map was reprojected so that the central meridian is located at 180$^\circ$ longitude. Panel b shows the magnetic field polarity at the source surface ($r$ = 2.0 R$_\odot$), showing positive (blue) magnetic polarity in the northern hemisphere and negative (red) polarity in the southern hemisphere separated by the heliospheric current sheet (HCS; blue line). The footpoints and their polarity of open magnetic field lines are shown in panel c, and the extrapolated field lines are illustrated on top of the AIA 193 \AA\ image in panel d. 

As this observation period is close to solar minimum, the polar regions are dominated by open-field footpoints that manifest themselves as polar CHs. The targeted low-latitude CH corresponds to the positive open-field regions located at 180$^\circ$ longitude. This CH extends beyond the equator into the southern hemisphere, resulting in a warped HCS that runs adjacent to the leading, trailing, and bottom boundaries of the CH, as shown in panel b of Figure~\ref{fig:PFSS_summary}. Hence, these regions adjacent to three sides of the CH should correspond to helmet streamers, with the leading and bottom boundaries located closer to the streamers than the trailing boundary. Meanwhile, the north section of the CH corresponds to a pseudostreamer, which also connects to the northern polar CH. The approximate locations of the helmet streamers and a pseudostreamer are indicated in panel b of Figure~\ref{fig:PFSS_summary}.

% Discuss comparison between PFSS footpoint and the obtained CH boundary.
The location of open-field footpoints, as shown in panels c and d of Figure~\ref{fig:PFSS_summary}, seems to generally agree with the CH area identified using AIA 193 \AA\ images, although some mismatches are still evident. 
% However, some mismatches are still evident, such as the presence of open-field footpoints outside the northwest CH boundary, or the lack of open fields near the bottom boundary. 
It has been demonstrated that the open-field regions derived from coronal modelling can differ considerably from CH areas derived from EUV observations \citep[e.g.,][]{Heinemann2023, Asvestari2024, Heinemann2026}, highlighting the limitation of current modelling in accurately representing the solar corona. Nevertheless, the PFSS model result shown here should still be useful as a simple representation of the magnetic field configuration of the targeted CH and surrounding regions.

\section{Results} \label{S-res}
\subsection{Differential Emission Measure Analysis} \label{subsec:DEM}
Assuming that EUV emissions in the solar corona are emitted from optically thin plasma in thermal and ionisation equilibrium, the amount of emitting plasma in a small volume can be described by emission measure (EM). For each column covering a small region on the solar disc (corresponding to a pixel in spatially resolved observations) with height $h$, the EM is given by,
\begin{equation}
    \text{EM} = \int N^2_e dh.
\end{equation}
In the case of multi-thermal plasma, the temperature distribution of the plasma with column height $h$ is described by the differential emission measure (DEM) function $\phi(T)$, given by
\begin{equation}
    \phi(T) = \frac{d~\text{EM}}{dT} = N^2_e \frac{dh}{dT}.
\end{equation}

%consider removing the density results? (seems fine for now)
Consequently, DEM-weighted averaged electron temperature $\bar{T}$ and averaged electron density of the plasma along the LOS $N_e$ are then given by \citep[e.g.,][]{Cheng2012, Long2019, Saqri2020},
\begin{equation}
    \bar{T} = \frac{\int \phi(T) ~T~ dT}{\int \phi(T)~dT} = \frac{\int \phi(T) ~T~ dT}{EM},
\end{equation}
and, 
\begin{equation}
    N_e = \sqrt{\frac{EM}{h}},
    \label{eq:ne_DEM}
\end{equation}
where $h$ is the LOS integration height. In this analysis, $h$ is approximated to be the hydrostatic scale height defined as 
\begin{equation}
    h = \frac{k_B\bar{T}}{mg\mu},
\end{equation}
where $k_B$ is the Boltzmann constant, $\bar{T}$ is the EM-weighted temperature, $m = 1.67\times10^{-27}$ kg is proton mass, $g = 274~\text{m}~\text{s}^{-2}$ is the Sun's gravitational constant, and $\mu = 0.6$ is the mean molecular weight assuming fully ionised plasma. Note that the hydrostatic scale height may not accurately represent the actual column height of emitted plasma, as the CH boundary regions contain a mixture of open and closed magnetic field configurations.

The DEM function $\phi(T)$ of the multi-thermal coronal plasma can be obtained from the EUV emissions observed by AIA. The AIA observables in the EUV passband can be described as a set of linear equations \citep{Hannah2012, Hannah2013} as follows:
\begin{equation}
    g_i = \textbf{K}_{i,j}~\phi(T_j) + \delta g_i
    \label{eq:DEMinversion_discrete}
\end{equation}
where $g_i ~(i = 1,..., M)$ is the observed intensity of each passband with observational uncertainties $\delta g_i$, $T_j ~( j = 1,..., N)$ denotes the specific temperatures of the DEM function, and $\textbf{K}_{i,j}$ is the kernel corresponding to the temperature response function of each AIA passband \citep{Boerner2012}. The temperature response function is derived using the emissivity model from the CHIANTI atomic database version 9.3 \citep{Dere1997, Dere2019}, assuming coronal abundance. Note that the plasma composition at the CH boundary regions is likely to be in the middle between photospheric and coronal abundance \citep{Koukras2025}. 

The DEM function in each temperature bin $\phi(T_j)$ can then be derived by solving the inversion problem in Equation \ref{eq:DEMinversion_discrete}. In this analysis, we employ the regularised inversion method developed by \citet{Hannah2012, Hannah2013} to calculate the DEM of the plasma at the CH boundary. The prepared AIA data in six EUV passbands ($M=6$): 94~\AA, 131~\AA, 171~\AA, 193~\AA, 211~\AA, and 335~\AA , are used as input. The instrumental uncertainties, the uncertainties of the atomic model, and the mean value of the residual light observed during lunar eclipses \citep{Heinemann2021} are taken into account to estimate the error of each pixel in the input EUV images. The analysis is performed over the temperature ranges of log($T$) = 5.5 - -6.5 ($T$ = 0.3 -- 3 MK), equally spaced logarithmically into 20 temperature bins ($N=20$). We chose this temperature range to avoid artefacts in DEM analysis from low count rates in some passbands \citep{Hofmeister2025}.

\begin{figure*}[t!]
    \centering
    \includegraphics[width = \textwidth]{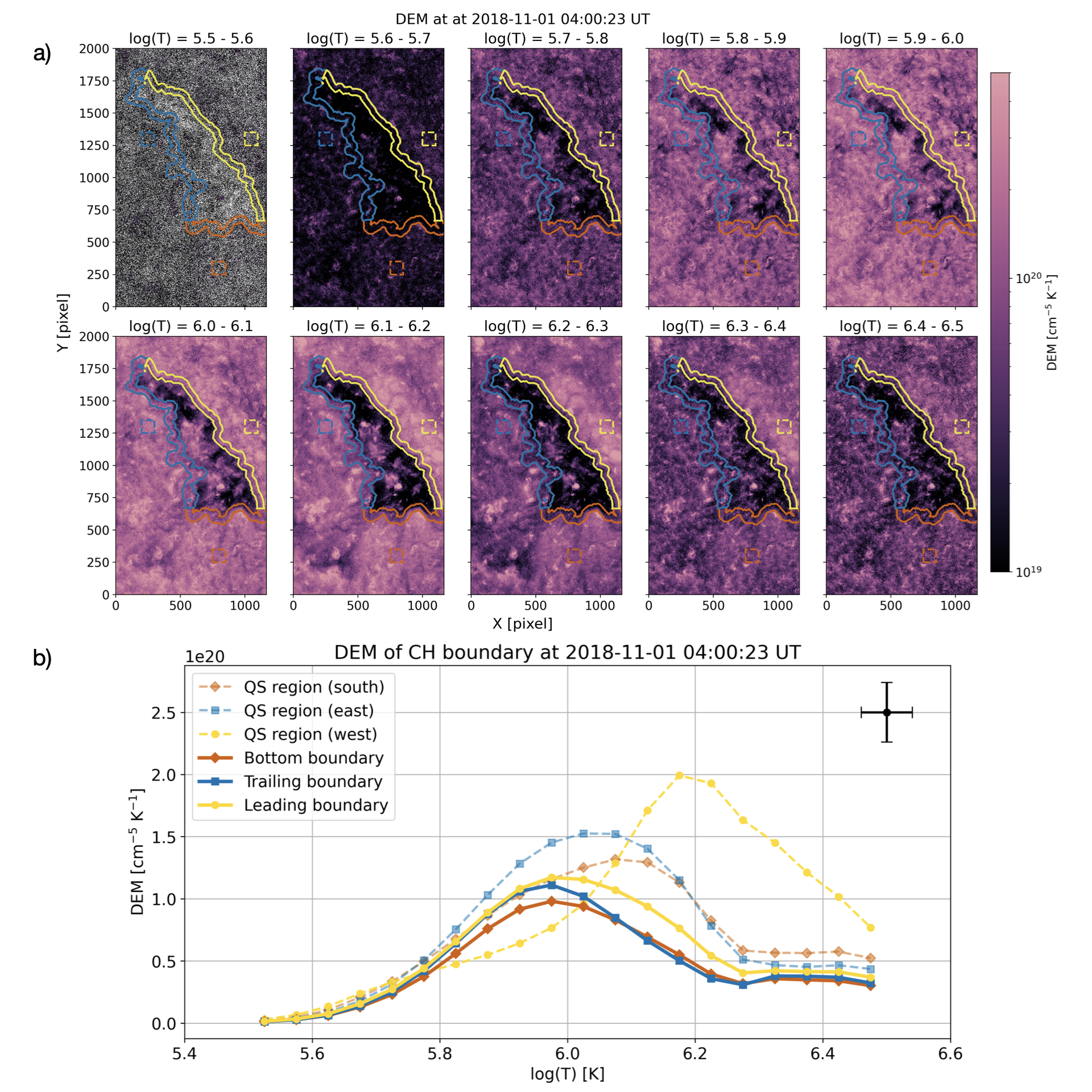}
    \caption{DEM of the CH plasma on 2018 November 1 at 04:00 UT. (a) The spatially resolved DEM distribution of the CH and surrounding regions in 10 temperature bins. The red, blue and yellow contours indicate the bottom, trailing and leading boundary regions of the CH. The dashed-line squares indicate the sample of quiet Sun regions in south (red), east (blue) and west (yellow) direction from the CH. (b) Average DEM distributions of the bottom, trailing, and leading boundary of the CH. The average DEM distributions of quiet Sun regions are also plotted in coloured dashed lines. The black errorbar indicates the representative uncertainty of DEM.}
    \label{fig:DEM_1day}
\end{figure*}

Figure \ref{fig:DEM_1day} displays the results of the DEM analysis using the AIA observations at 04:00 UT. Panel a shows the DEM map of the CH in each temperature bin. Note that there are several data points for which the DEM inversion method cannot compute valid results in a reasonable amount of time. The invalid data points scatter uniformly throughout the images, as shown by the white pixels in the upper left of panel a. For the remaining subpanels, we interpolated the invalid data points for illustrative purposes only. These data points are then excluded from further analysis.

The CH can be clearly identified at log($T$) = 6.1 -- 6.2 and log($T$) = 6.1 -- 6.3, where the DEM values in the CH are significantly lower than those of the surrounding regions. The DEM distributions of QS regions adjacent to each CH boundary side also show a noticeable difference. In particular, the DEM at log($T$) = 6.1 -- 6.3 of the QS region at the western side of CH is higher than other sides, resulting in a visually clear contrast between the CH and non-CH plasma and a well-defined visual boundary at those temperature ranges. 

To highlight the difference in the DEM distributions in each part of the CH boundary, we define the CH boundary region as regions within 25 pixels of the derived CH boundary line on each side, resulting in the band of $\sim 21$ Mm width, with the CH boundary at the centre of the band. The width is chosen to correspond approximately to the characteristic scale of the CH boundaries of $\sim 15 - 30$ Mm \citep[cf.][]{Heinemann2021, Koukras2025}, as well as to the scale on which the irregularities of the CH boundary are computed in Section \ref{subsec:CDM}. Then we further divided this boundary region into three sections: bottom, trailing, and leading boundaries. Each section is highlighted in colour contours in panel a of Figure \ref{fig:DEM_1day} (red: bottom, blue: trailing, yellow: leading), and the comparison of the average DEM distribution of three CH boundary sections is shown in panel b. The representative uncertainty in the derived DEM is indicated by a black error bar. Note that the uncertainty in the DEM axis is dominated by systematic errors in the atomic model that are used to derive the temperature response function, while the error in the temperature axis is related to the temperature resolution of the DEM fitting \citep{Hannah2012}.

The DEM distributions peak at log($T$) = 5.9 -- 6.0 ($T$ $\sim$ 0.8 -- 1.0 MK) for all CH boundary sections, which is consistent with the peak temperature of the CH plasma \citep{Heinemann2021}. The DEM peak of the bottom boundary is slightly lower than that of the leading and trailing counterpart. The difference in the DEM value is also noticeable at log($T$) = 6.0 -- 6.2, indicating that there is more plasma with log($T$) $> 6.0$ at the leading boundary compared to the trailing and bottom boundaries. While the DEM for the trailing boundary drops significantly at log($T$) $>$ 6.0, the DEM curve of the leading boundary has a broader peak and decreases less rapidly with temperature. Interestingly, the shape of the DEM distribution of the bottom boundary is similar to that of the leading boundary, although the actual DEM values are lower in the range log($T$) = 5.8 -- 6.3.

To confirm that the difference in the DEM distribution is not an artefact of the DEM calculation, we also checked the average EUV intensities of the leading, bottom, and trailing boundaries. The results are shown in Figure \ref{fig:AIA_CHBR} in Appendix~\ref{S-A2}. The average intensity of the 193 \AA\ passband (characteristic log($T$) = 6.2) is noticeably lower in the trailing boundary than in the leading boundary, which is consistent with our DEM calculation.

% add quiet Sun discussion
The average DEMs of the QS regions (east, west, and south of the CH) are also shown in panel b of Figure \ref{fig:DEM_1day} as comparisons. Overall, the DEMs of the QS regions peak at a higher temperature than those of the CH boundaries, and the peak temperature and distribution shape are comparable to previous work \citep{Brooks2009, Milanovic2025}. The DEM of the east QS region peaks at a temperature of log($T$) = 6.0-6.1. For the south QS region, the DEM peaks at a similar temperature, although the distribution is skewed towards higher temperatures. The west QS region has a clear peak in DEM at log($T$) $\sim 6.2$ and a significantly different DEM distribution compared to the east and south  QS regions, indicating that the west QS region has more hot plasma than others. 

\subsection{Magnetic Field Properties} \label{subsec:mag}
% New table (25pixel annulus)
\begin{table}[t!]
    \begin{tabularx}{\textwidth}{@{}lccc@{}}
    \toprule
    Sections  & $\bar{B}$ (G) & $\bar{B}_{us}$ (G) & Flux Imbalance Ratio \\ \midrule
    Bottom boundary   & 1.4       & 7.1       & 0.19                             \\
    Trailing boundary & 1.2    & 7.5      & 0.16                            \\
    Leading boundary  & 3.0       & 8.4       & 0.36                           \\ \bottomrule
    \end{tabularx}%
\caption{Magnetic field properties of each CH boundary section (as indicated by colour contours in Figure \ref{fig:DEM_1day}) on 2018 November 1, 04:00 UT. $\bar{B}$ and $\bar{B}_{us}$ refer to signed and unsigned magnetic flux density, respectively. The flux imbalance ratio corresponds to the ratio between signed and unsigned magnetic flux densities.}
\label{tab:HMIresult_overview}
\end{table}

The evolution of the magnetic field in the CH can be investigated by calculating several magnetic field properties derived from HMI observations. For each pixel of the HMI magnetogram, the magnetic field strength and the projected area on the disc must be corrected for projection effects \citep{Hofmeister2017}. By assuming that the magnetic field is radial, the corrections are
\begin{align}
    B_i &= \frac{B_{i, LOS}}{cos\alpha_i},\\
    A_i &= \frac{A_{i, proj}}{cos\alpha_{i}},
\end{align}
where $B_i, A_i$ is the projection-corrected magnetic field strength and the solar disc area for each pixel, $B_{i, LOS}$ is the LOS magnetic field strength, $A_{i, proj}$ is the pixel area of 0.505\SI{}{\arcsecond}$\times0.505$\SI{}{\arcsecond}, and $\alpha_i$ is the angle from the solar disc centre to each pixel. 

Next, the signed magnetic flux $\Phi$, the unsigned magnetic flux $\Phi_{us}$, the signed magnetic flux density $\bar{B}$, and the unsigned magnetic flux density $\bar{B}_{us}$ can be calculated as
\begin{align}
    \Phi &= \sum_i B_{i}A_{i} \\
    \Phi_{us} &= \sum_i |B_{i}A_{i}| \\
    \bar{B} &= \frac{\Phi}{\sum_i A_{i}} \\
    \bar{B}_{us} &= \frac{\Phi_{us}}{\sum_i A_{i}}.
\end{align}
Additionally, we also computed the magnetic flux imbalance ratio from the ratio of $\Phi$ and $\Phi_{us}$.

% Add something on the table 1.
The magnetic field properties of the CH boundary are shown in Table \ref{tab:HMIresult_overview}. Using the same segmentation described in Section \ref{subsec:DEM}, we find noticeable differences in the magnetic characteristics of the bottom, leading, and trailing boundary of this CH. In particular, $\bar{B}$ of the leading boundary is almost 2--3 times higher than the trailing and bottom boundaries, although $\bar{B}_{us}$ are relatively more similar in three sections. This is further highlighted in the flux imbalance ratio reported in the fourth column, which can be used as a proxy for magnetic unipolarity. Using this proxy, we can conclude that the leading boundary is distinctly more unipolar than the trailing and bottom boundaries.

\subsection{Irregularities of Coronal Hole Boundary}\label{subsec:CDM}

\begin{figure*}[t!]
    \centering
    \includegraphics[width = \textwidth]{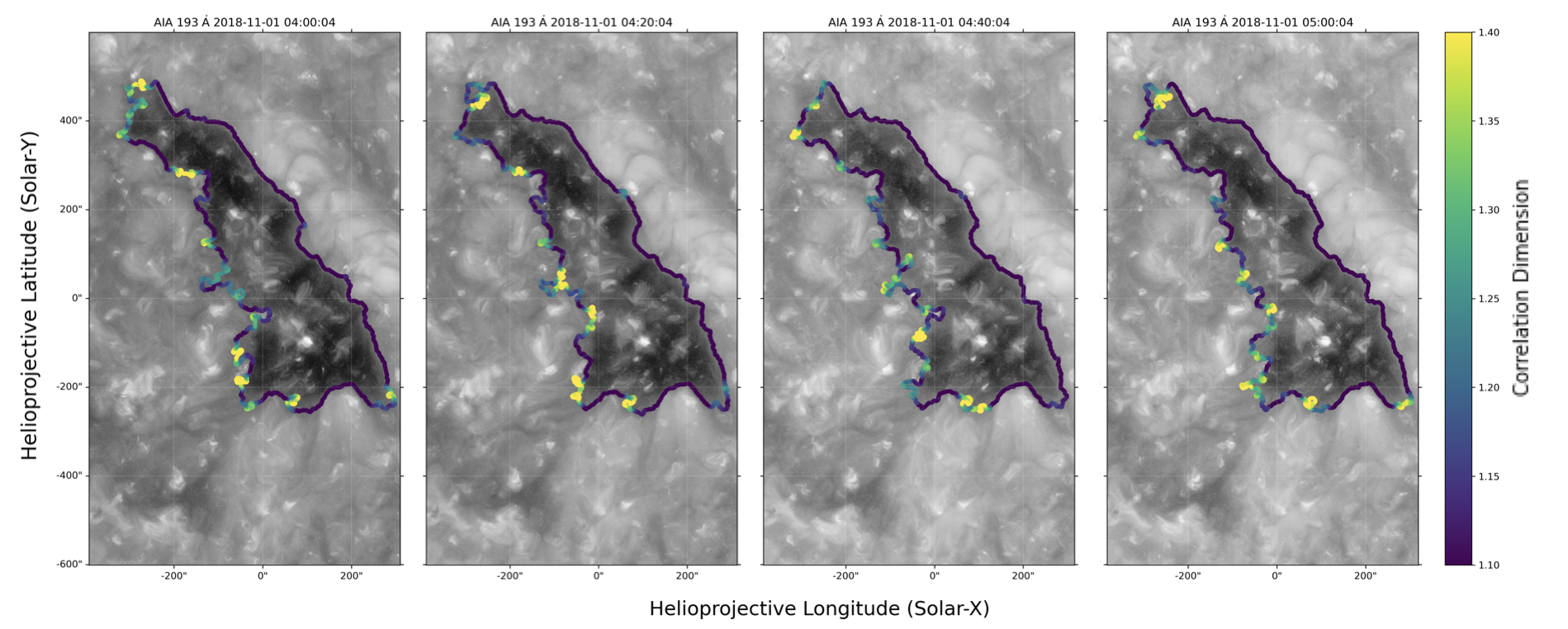}
    \caption{Correlation dimension map overlaid on the boundary of the CH across a 1-hour period, from 2018 November 01 04:00 to 05:00 UT. A 30 s animation of this figure is available as electronic supplementary material, covering an hour of solar time.}
    \label{fig:CDM_1hour}
\end{figure*}

The irregularities of the CH boundary can be quantified using the Correlation Dimension Mapping (CDM) method developed by \citet{Mason2022}. This method calculates a parameter called the correlation dimension $D$ for each point along the boundary, which indicates the local irregularity of the boundary line. Although designed for CH boundary characterisation, this method can be adapted to wider solar physics applications, such as analysing the fine structures of flare ribbons \citep{CorchadoAlbelo2026}.

In the two-dimensional plane, the correlation integral $C(r)$ is numerically approximated to be
\begin{equation}
    C(r,x,y) = \sum_{i=1}^{N(x,y)}\Theta(r-\sqrt{(x_i-x)^2 - (y_i-y)^2}),
    \label{eq:CDM}
\end{equation}
where ($x_i,y_i$) are the coordinates of the data points surrounding the reference location ($x,y$) and $\Theta$ is the Heaviside step function. 

For a self-similar set of data points %(i.e., a fractals-like structure) 
and small $r$, $C(r,x,y)$ takes the power law form,
\begin{equation}
    C(r,x,y) \sim r^{D(x,y)},
    \label{eq:Dfactor}
\end{equation}
where $D(x,y)$ is the locally defined correlation dimension that characterises the geometry of the data points. In general, $D = 1$ corresponds to a straight line with minimal complexity, and a more complex and irregular boundary corresponds to a higher $D$ value.

To analyse the complexity of the CH boundary, $D$ is calculated in the radius range $r$ of 6--25 pixels that surround each CH boundary pixel, equivalent to scales of 5--20 Mm. This radius range is chosen to be in line with previous CDM analysis on the CH boundary by \citet{Mason2022}. The lower end of the scale corresponds to the sizes of small-scale transient events (e.g., jetlets), while the upper end corresponds to supergranulation scales. 

Figure \ref{fig:CDM_1hour} and its animated version show the evolution of the CH boundary and its irregularity determined from CDM. During the 1-hour observation period, there appear to be no large-scale changes of the CH boundary. It is evident from the distribution of the $D$ values that each part of the CH boundary exhibits different degrees of irregularity, with the high $D$ values spatially localised at specific parts along the boundary. In particular, the leading boundary in the north-west seems to have a much smoother and less complex boundary compared to the trailing boundary in the south-east. The overall values of $D$ also do not show a clear evolution with time, i.e., the more complex parts of the boundary remain highly complex throughout the observation period and vice versa. 

\subsection{Spatial and Temporal Variation of Coronal Hole Boundary Properties} \label{subsec:TDplot}
\begin{figure*}[t!]
    \centering
    \includegraphics[width=0.75\textwidth]{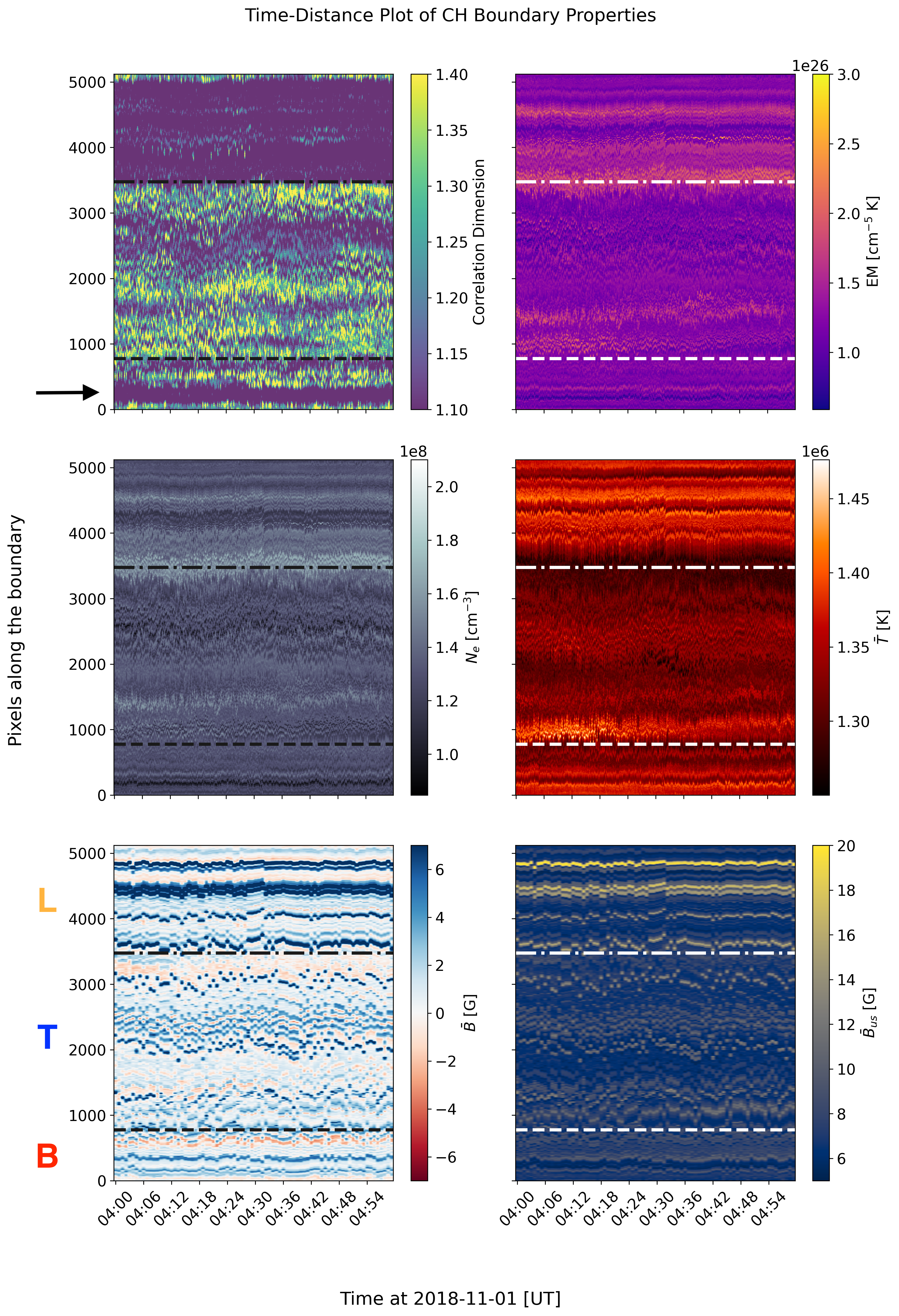}
    \caption{Time-distance stack plots of the various properties along the CH boundary, starting from the southwest corner of the CH. From top to bottom, the left column shows the correlation dimension $D$, EM-weighted temperature $\bar{T}$ and the signed magnetic flux $\bar{B}$, while the right column shows the EM, the electron density $N_e$, and the unsigned magnetic flux $\bar{B}_{us}$. Each section is separated by the dash-dotted lines. The leading, trailing and bottom boundaries are labelled as L, T, and B.}
    \label{fig:Time-distance-CHB}
\end{figure*}

The time-distance plots in Figure \ref{fig:Time-distance-CHB} visualise the spatial and temporal variation of the CH boundary properties over a period of 1 hour, including the correlation dimension $D$, plasma properties (EM, $\bar{T}$, $N_e$) and magnetic field properties ($\bar{B}$ and $\bar{B}_{us}$). $D$ values are calculated for each pixel along the boundary, while the plasma and magnetic field properties are the average values in a 50-pixel$\times$50-pixel box surrounding each boundary pixel, corresponding to the upper scale of $D$ calculation ($r$ = 25 pixels). 

\begin{figure*}[t!]
    \centering
    \includegraphics[width=\linewidth]{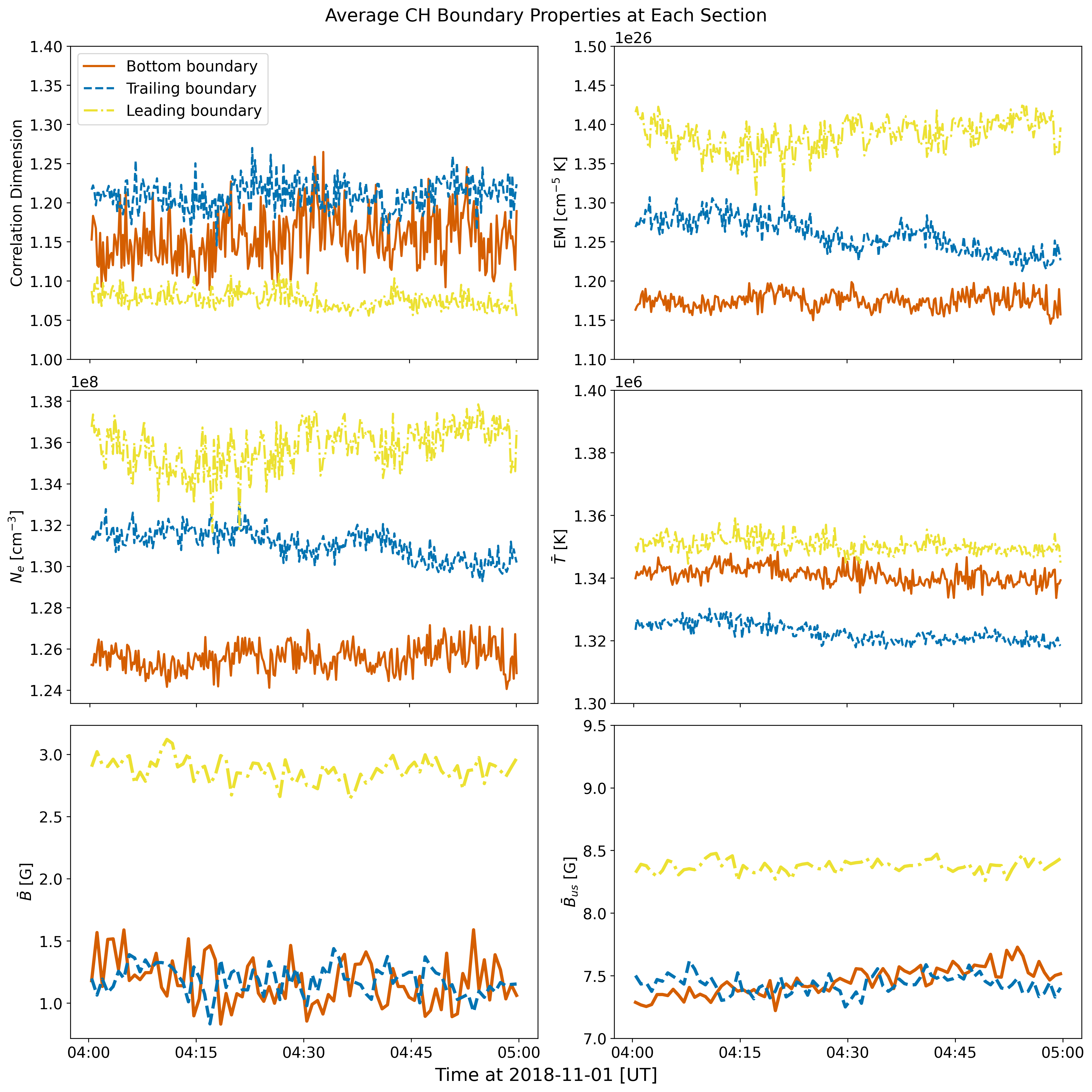}
    \caption{The average correlation dimension, plasma parameters and magnetic field properties inside each coronal hole boundary section from 04:00 UT to 05:00 UT. Similar to Figure~\ref{fig:Time-distance-CHB}, the left column shows the correlation dimension $D$, the EM-weighted temperature $\bar{T}$ and the signed magnetic flux $\bar{B}$, while the right column shows the EM, the electron density $N_e$, and the unsigned magnetic flux $\bar{B}_{us}$.}
    \label{fig:CHB-prop-mean}
\end{figure*}

The starting pixel corresponds to the bottom-right corner of the CH boundary, counting in the clockwise direction. Since the numbers of boundary pixels fluctuate at each time step as the CH boundary length changes, the lengths of the boundary line are normalised to the mean CH boundary length during this observation period using a linear interpolation method that still preserves the evolutionary trend of the data \citep{Mason2022}. Similar to Section \ref{subsec:DEM}, the normalised CH boundary line is then divided into three sections (leading, trailing, bottom) using the corners indicated by coloured dashed lines in Figure \ref{fig:Time-distance-CHB}. We also calculate the average value of each of the CH boundary properties in each section and plot their variations with time in Figure \ref{fig:CHB-prop-mean}.

Although there is no clear temporal evolutionary trend of the CH boundary properties that can be observed during this 1-hour period, the differences between the properties of each section of the CH boundary are evident. As shown in Figure \ref{fig:CHB-prop-mean}, the leading boundary appears to have a higher $\bar{T}$, $\bar{B}$, and $\bar{B}_{us}$ and a lower $D$ compared to other regions. The trailing boundary, on the other hand, is generally much more complex and lower $\bar{T}$. The time-distance plot of $\bar{B}$ (see bottom left panel of Figure \ref{fig:Time-distance-CHB}) also shows that the trailing boundary appears to have a more negative flux, resulting in the region becoming relatively less unipolar. 

Interestingly, the bottom boundary shows characteristics of both the leading and trailing boundaries. Looking at the time-distance map in Figure \ref{fig:Time-distance-CHB}, it appears that the western half of the bottom boundary (around pixel numbers 100--500; see the black arrow in Figure \ref{fig:Time-distance-CHB}) has the characteristics of the leading boundary, including a smoother boundary, relatively high temperature, and a more unipolar magnetic field. Meanwhile, the other half of the bottom boundary (around pixel numbers 500--800) appears to be  more spatially irregular, less unipolar, and has slightly lower temperature.

% Add summary
Overall, this time-distance analysis highlights the difference in spatial irregularity at each of the CH boundary sections, while the differences in plasma and magnetic field properties are still also evident. Comparison of the leading and trailing boundary of the CH also hints at a possible connection between boundary complexity, temperature, and magnetic field unipolarity.

\section{Summary and Discussion} \label{S-disc}
In this paper, the properties of the boundary of a low-latitude CH during the 1-hour observation period from 2018 November 1, 04:00 UT to 05:00 UT  are derived using AIA and HMI observations. The large-scale magnetic configuration in the corona is derived using PFSS extrapolation to provide a context of the magnetic environment of the CH and surrounding regions. The DEM analysis is used to derive the plasma properties from multipassband EUV observations, while the irregularities of the boundary line are derived from the CDM method.

We found that different parts of the CH boundary exhibit different plasma and magnetic field properties and that these differences are present for most of the one-hour observation period. The most obvious difference is seen in the DEM distribution shown in Figure \ref{fig:DEM_1day}, where the leading and bottom boundaries have relatively more plasma at temperatures higher than 1 MK compared to the trailing boundary. This is also reflected in a slightly higher averaged EM-weighted temperature in these regions, although the difference is less pronounced. Moreover, the leading and bottom boundaries also have higher unbalanced and unsigned magnetic flux densities, indicating that they have a stronger magnetic field and are more unipolar than the trailing boundaries. 

% {\bf --- Move this paragraph up from previous version}
The difference in the DEM distribution and the DEM-weighted temperature at the leading and trailing boundaries may hint towards the different plasma heating, possibly caused by the nature of interchange reconnection. The numerical simulation \citep[e.g.,][]{Wang1996, Lionello2005, Mason2023} shows that interchange reconnection at the CH boundary can be driven by differential rotation. The open field lines are constantly closed down at the leading (western) boundary, while the closed field lines are opened up at the trailing (eastern) boundary. This leads to more open flux concentrated at the trailing boundary, as seen in the time-dependent MHD simulation of \citet{Mason2023}, and may help explain the lower temperature of the trailing boundary that we observed for this CH. Moreover, the energy flux density from interchange reconnection also depends on the magnetic field strength at the footpoints \citep{Wang2020, Wang2024}. Hence, the higher temperature of the leading boundary may also be due to the stronger magnetic field compared to the trailing boundary. However, a statistical study including several CHs across a solar cycle is needed to confirm whether all CH leading boundaries are hotter or more unipolar than the trailing counterpart and whether this phenomenon is mainly driven by opening/closing of field lines. This is beyond the scope of the current work, although it would be an interesting topic to investigate further.

It can also be observed that different parts of the CH boundary line have different degrees of spatial complexity, as shown in Figure \ref{fig:CDM_1hour}. In particular, the leading boundary is relatively smooth, in contrast to the trailing boundary, which is much more complex. The different characteristics of plasma and magnetic field are also evident, highlighting the different nature of each part of the CH boundary and suggesting a possible connection between boundary irregularity and local plasma conditions and magnetic field environments.

% Discuss the relative stability of boundaries, refer to results by Ugarte-Urra 2026
During the 1-hour observation period, we also find that there are no clear, systematic evolutionary trends in the boundary shape and the CH boundary properties. \citet{Mason2022} reported similar results that the CH boundary generally exhibits near steady-state conditions on an hour time scale, with the exception of sporadic jetting events near the boundary. \citet{Ugarte-Urra2026} proposed that the evolution of the CH boundaries in the absence of nearby ARs is mainly due to the supergranular convection that advects the open field lines and drives the interchange reconnection with long, closed loops. This process, however, operates on a timescale of 1--2 days \citep[i.e., lifetime of supergranules;][]{Rincon2018}, which may also partly explain why we did not observe significant changes in the CH boundaries over an hour period.

\citet{Kahler2002} identified three types of CH boundary: diffuse-field boundary, matching-polarity AR boundary, and opposite-polarity AR boundary. They found that matching-polarity AR boundaries are generally smooth and sharply defined, while the diffuse-field boundaries are more ragged and non-uniform. Similar results are also reported by \citet{Reiss2024}, who notice that the existence of filaments and/or ARs near the CH boundaries results in a distinctively different appearance of the boundary lines compared to those bordering the QS region. In our case, the leading boundary generally appears to be smoother, while the trailing boundary appears to be more irregular. Hence, the leading boundary may correspond to the matching-polarity AR boundary reported by \citet{Kahler2002}. In particular, a decayed NOAA AR 12317 was located west of this CH during the Carrington rotation 2207, three Carrington rotations earlier than the observation period. The decay of AR also results in the formation of the filament channel west of the CH, as shown in Figure \ref{fig:AIA_HMI}. Hence, the region outside the leading boundary corresponds to decayed AR loops that overarch the filament channel, which is a relatively coherent and organised magnetic field structure and may lead to a smoother boundary line. Meanwhile, the trailing boundary corresponds to the diffuse-field boundary, in which case its irregularity might reflect the underlying network magnetic field \citep{Kahler2002}.

The regions between the leading boundary and the filament channel have cellular configurations (shown by the red arrow in Figure \ref{fig:AIA_HMI}), identical to the coronal cell structure observed by \citet{Sheeley2012}. The authors found that the centres of these coronal cells are located at the strong photospheric magnetic elements, and the cells appear as plume-like elongated structures when observed near the limb. The arrangement of these cells is also related to the direction of the horizontal magnetic field of the filament channel, analogous to the chromospheric fibrils \citep{Sheeley2013}. Coronal cells are interpreted as the footpoints of extended closed loops that arch over the filament channel and connect to distant regions of opposite polarity. The characteristics of coronal cells are evident in this case, and the interpretation is also compatible with the global magnetic field of this CH derived from PFSS extrapolation (Figure \ref{fig:PFSS_summary}). Note that \citet{Alzate2025} found similar coronal cell structures inside the CH, which also correspond to plume-like structures in open-field regions. One main difference is that cells inside CH can only be observed using AIA 171 {\AA}.

\begin{figure*}[t!]
    \centering
    \includegraphics[width=0.8\textwidth]{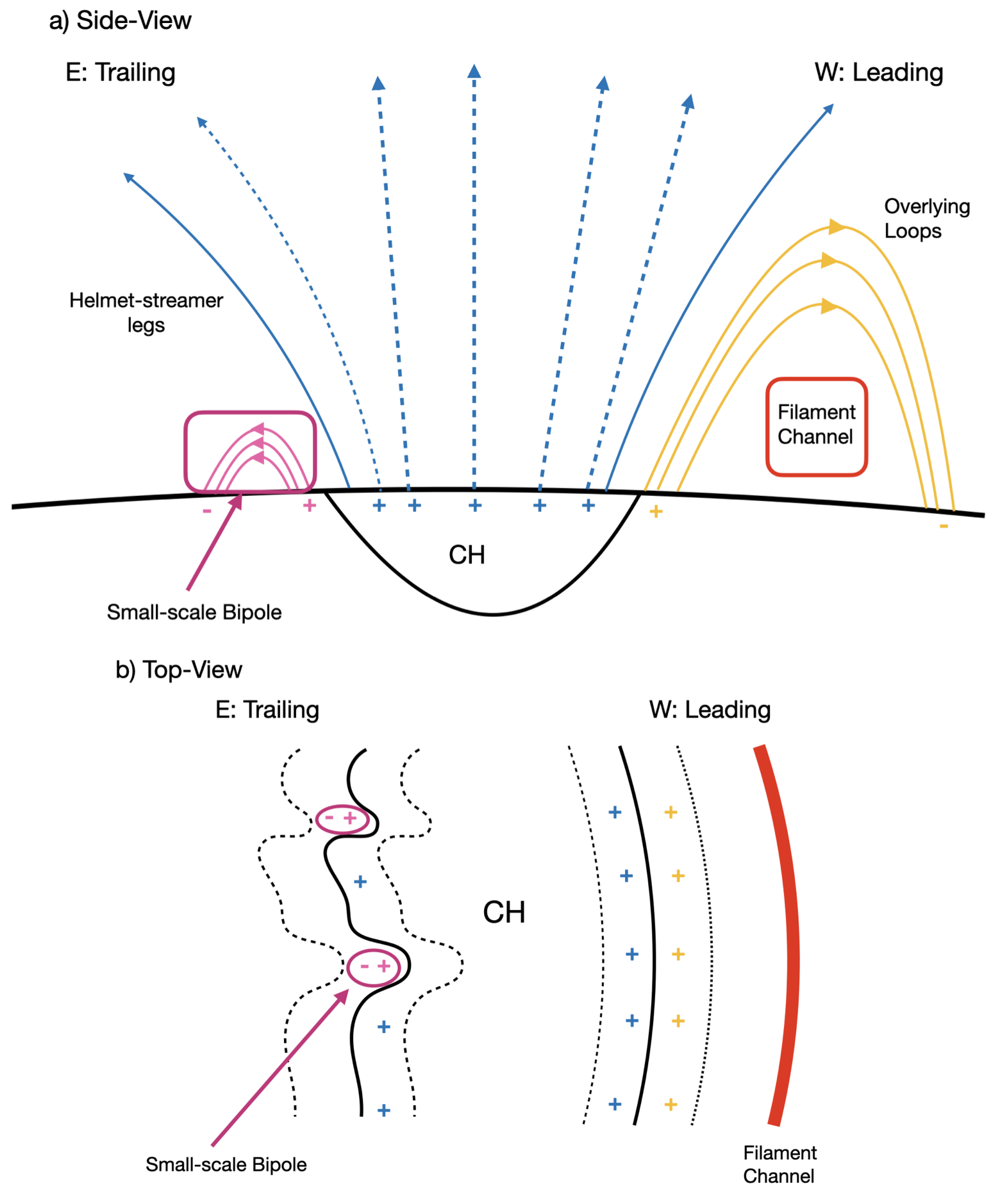}
    \caption{
    Simplified diagrams illustrating the overall structure of the leading (western: W) and trailing (eastern: E) boundary of the observed CH from a side view (panel a) and top view (panel b). The blue field lines are the CH field lines: dashed lines are open field lines, and the solid lines are closed field lines making up the helmet streamer legs. The leading boundary consists of large well-organised coronal loops overaching the filament channel, denoted by yellow lines. Meanwhile, the trailing boundary consists of randomly orientated, small magnetic bipoles, indicated by pink lines. The solid and dashed black lines in panel b indicate the CH boundary lines and CH boundary regions.}
    \label{fig:CHBdiagram}
\end{figure*}

All of these results strongly suggest that the differences in the properties of the CH boundaries are primarily due to the magnetic configuration of the adjacent regions. In particular, the difference in the leading and trailing boundaries of the observed CH can be described by the simplified diagrams in Figure \ref{fig:CHBdiagram}.

The leading boundary is located close to the filament channel. Several large loops, likely from the decayed AR, arch over the filament channel and give the appearance of coronal cells. The strong magnetic elements at the loop footpoints may be partly responsible for the strong heating and higher temperature. These overlying loops therefore define the smoothness of the leading boundary for two reasons. The first reason is the sharp temperature gradient between open and closed field structures (see panel a of Figure \ref{fig:DEM_1day}). The second reason is that the arrangement of loop footpoints is rather structured along the direction of the filament channel, as evidenced by the leading boundary being parallel to the filament channel.

On the other hand, the complex trailing boundary can be explained by the more complex mixed polarity field.
The presence of opposite-polarity flux creates numerous small-scale bipoles. The open field line footpoints then have to be located around the pockets of closed magnetic loops, naturally leading to the more corrugated CH boundary lines, as shown in panel b of Figure \ref{fig:CHBdiagram}.
AIA observations reveal that there are numerous small-scale bright points located near the boundary, all of which correspond to small magnetic bipoles. The appearance of these bright points and their rather random orientation may be responsible for the irregular nature of the boundary line. Although these bright points can cause localised increases in plasma temperature, density, and magnetic field strengths, the overall magnetic field is more dispersed and there is a lack of organised structure (like a large-scale loop system above a filament channel) near the trailing boundary, both of which may result in a lower average temperature.

\citet{Aslanyan2022} and \cite{Mason2022} proposed that the boundary between a CH and a pseudostreamer is generally smoother than the boundary between a CH and a helmet streamer because interchange reconnection occurs more readily in the pseudostreamer.
In our case, Figure \ref{fig:PFSS_summary} shows that the leading, trailing, and bottom boundaries of this CH all border the helmet streamer, with only the small tip at the top bordering the pseudostreamer. Still, by applying the same CDM method, we observe striking differences in the correlation dimension of each boundary line. This suggests that the CH boundary irregularity does not depend solely on the nature of reconnection dynamics (such as helmet streamer vs pseudostreamer reconnection) but also depends on local magnetic structures, such as large, organised loops over the filament channel or low-lying coronal bright points. 

 % {\bf --- Add paragraph on the typical/atypical CH} 
% Add paragraph on how it could be repeated for other CH
This work presents a case study of the possible connection between spatial irregularities and the physical properties of the boundary regions of a low-latitude CH. Although one CH may not be a representative of all CHs, we argue that many low-latitude CHs have characteristics similar to those of the CH we investigated in this work. Decayed ARs are usually related to the formation of low-latitude CHs \citep[e.g.,][]{Karachik2010, Petrie2013}, as well as the formation of filaments. Consequently, low-latitude CHs are usually located near decayed ARs and filaments, \citep[see Table 2 in][]{Reiss2024} and coronal cells should also be commonly found at the boundary of those CHs. Hence, we suspect that the CH analysed can be considered a typical low-latitude CH. However, further investigation should be done to determine whether the connection between spatial complexities and physical properties is also evident in other CHs. Again, a statistical study will help to confirm the scenario discussed in this paper.

Lastly, it is plausible that the solar wind originating from different CH boundaries may exhibit different properties. The CH studied in this work is magnetically connected to PSP during its first perihelion \citep[e.g.,][]{Badman2020}, and there are several works discussing the solar wind associated with this CH. In particular, \citet{Macneil2020} reported the enhancement of suprathermal electron flux associated with the slow-fast stream interface and proposed that the sources of these enhancements may be related to interchange reconnection at CH boundaries. \citet{Bercic2020} and \citet{Stansby2021} inferred the coronal electron temperature of this CH using in situ electron strahl measurements. They found that the inferred temperature is relatively low and uniform for solar wind streams originating from inside the CH. Meanwhile, solar wind streams from the CH leading boundary correspond to higher coronal temperatures. However, it should be noted that it is not clear how well the electron strahl/beams retain the coronal information in the presence of local transport processes \citep{Wu2026}.

%% Add results from Karna 2022 
The fast solar wind streams ($\sim$600 km s$^{-1}$) originating from this CH were also detected by the Advanced Composition Explorer \citep[ACE;][]{Stone1998} spacecraft at the L1 point and were analysed in detail by \citet{Karna2022}. Figure 7 in their article shows that the heavy-ion charge-state ratios (O$^{7+}$/O$^{6+}$, C$^{6+}$/C$^{5+}$) and elemental abundance ratios (Fe/O) decreased during the transition from slow to fast solar wind in the stream interaction regions. This decrease was consistent with the magnetic connectivity transition from helmet streamer to CH, passing through the leading boundary. Meanwhile, the change in elemental composition is less obvious at the trailing end of the fast solar wind streams. The relationship between the properties at the CH boundary and the solar wind emanating from them will be an interesting topic to explore in future work, especially as PSP approaches the perihelion at the heliocentric distance below 10~R$_{\odot}$ where the solar wind streams may better retain the information of the coronal source regions.

%%%%%%%%%%%%%%%%%%%%%%%%%%%%%%%%%%%%%%%%%%%%%%%%%%%%%%%%%%%%%%%%%%%%%%%%%%%
\begin{acks}
We would like to thank the reviewer for providing valuable suggestions and feedback that improve the manuscript. We are grateful to Huw Morgan, Colin Forsyth, Jesse Coburn, and Antoine Dolliou for their helpful discussions and suggestions. SDO data are courtesy of NASA/SDO and the AIA, EVE, and HMI science teams. This work utilises data produced collaboratively between the Air Force Research Laboratory (AFRL) \& the National Solar Observatory (NSO). The ADAPT model development is supported by AFRL. The input data utilised by ADAPT is obtained by NSO/NISP (NSO Integrated Synoptic Programme).
\end{acks}

\begin{authorcontribution}
N.N. led the project and data analysis, prepared all figures, and wrote the manuscript. D.M.L. and L.M.G. helped with project conceptualisation. S.G.H. developed the CATCH coronal hole detection algorithm. V.M.U. developed the CDM code. All authors contributed to the data analysis, discussion of the results and revision of the manuscript.
\end{authorcontribution}

\begin{fundinginformation}
N.N. was supported by STFC PhD studentship grant ST/W507891/1 and UCL Studentship during his PhD education at UCL. S.L.Y. would like to thank the Science Technology and Facilities Council for the award of an Ernest Rutherford Fellowship (ST/X003787/1). A.W.J. acknowledges funding from the STFC consolidated grant ST/W001004/1. S.G.H. acknowledges funding from the Austrian Science Fund (FWF) Erwin-Schr\"odinger fellowship [10.55776/J4560] and funding from the Research Council of Finland (Academy Fellowship) [370747; RIB-Wind].
\end{fundinginformation}

\begin{dataavailability}
SDO/AIA and SDO/HMI data can be accessed from the Joint Science Operations Centre (\url{http://jsoc.stanford.edu})
\end{dataavailability}

% \begin{materialsavailability}
% Information about available material ...
% \end{materialsavailability}

\begin{codeavailability}
This research made use of several open-source python packages including SunPy \citep{SunPyCommunity2020}, aiapy \citep{Barnes2020}, pfsspy \citep{Stansby2020b}, NumPy \citep{Harris2020}, Astropy \citep{AstropyCollaboration2013}, Matplotlib \citep{Hunter2007}, SciPy \citep{Virtanen2020},  scikit-image \citep{vanderWalt2014}, scikit-learn \citep{Pedregosa2011}, OpenCV \citep{Bradski2000}, asdf \citep{Greenfield2015}, seaborn \citep{Waskom2021}, cmcrameri \citep{Crameri2023}, and cupy \citep{Okuta2017}. This research used version 0.2.1 of the pyCATCH open-source software package \citep{Heinemann2019}, available at \url{https://github.com/sgheinemann/pycatch} The DEM analysis code in python is available as the demregpy package (\url{https://github.com/alasdairwilson/demregpy}) and at \url{https://github.com/ianan/demreg}. The CDM code was developed by V.M.U. and is available upon request.
\end{codeavailability}

\begin{ethics}
\begin{conflict}
The authors declare that they have no conflicts of interest.
\end{conflict}
\end{ethics}

% \noindent To change a title use an optional parameter:\par
% \verb+\begin{acks}[Acknowledgements]...\end{acks}+

%\acknowledgment US spelling: \verb+\acknowledgment+
%\acknowledgement British  spelling: \verb+\acknowledgement+

%%%%%%%%%%%%%%%%%%%%%%%%%%%%%%%%%%%%%%%%%%%%%%%%%%%%%%%%%%%%%%%%%%%%%%%%%%%
\appendix
%%% Add effect of varying the boundaries here.
\section{Comparison between Different Coronal Hole Boundary Thresholds} \label{S-A1}
\begin{figure}[ht!]
    \centering
    \includegraphics[width=\textwidth]{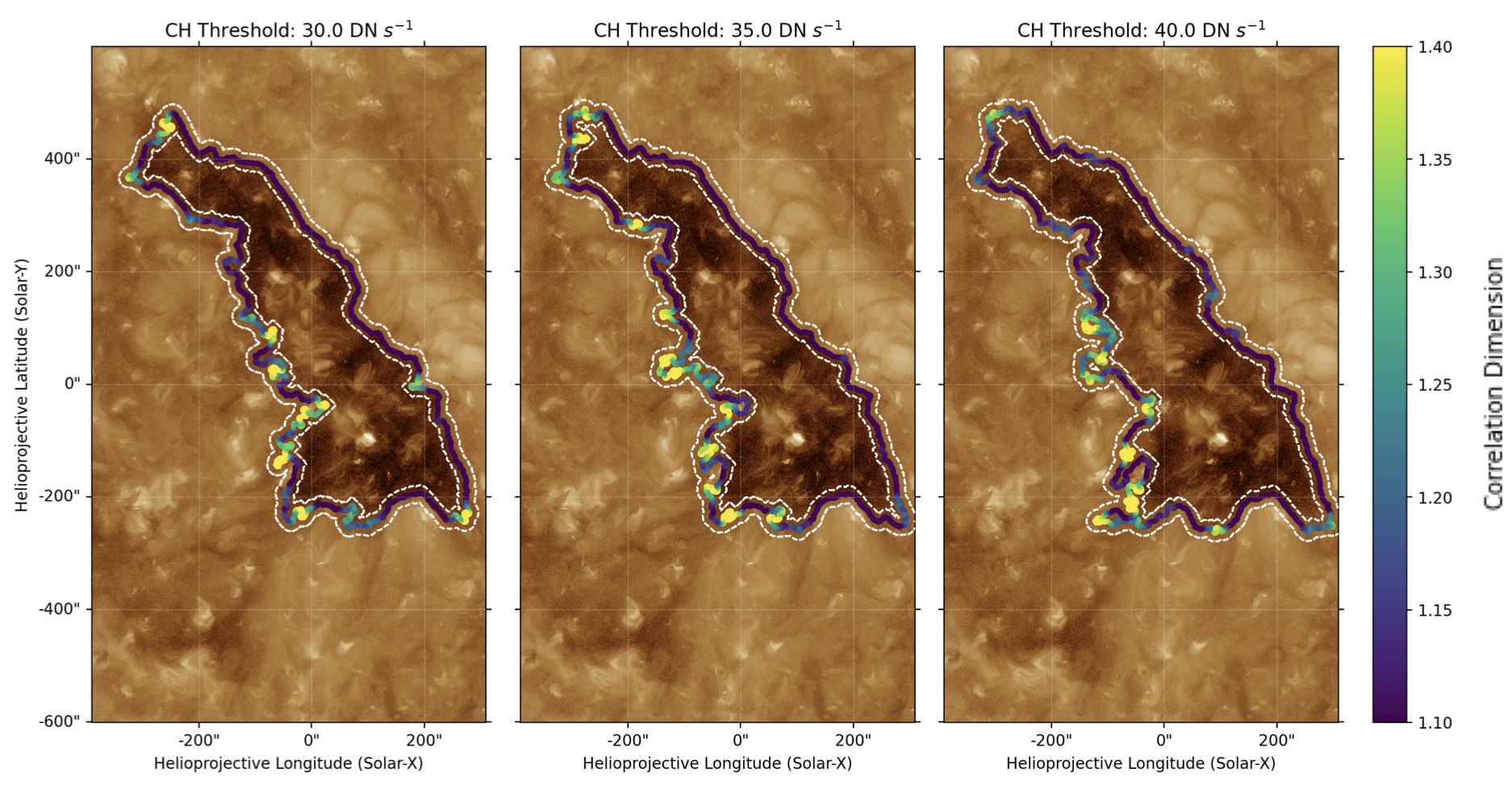}
    \caption{Comparison between CH boundaries and their associated spatial irregularities that are obtained using different intensity thresholds. The value of 35 DN~s$^{-1}$ (middle panel) is the value used in the analysis. The white dashed lines indicate the CH boundary regions.}
    \label{fig:AIA_CHbound_Compare}
\end{figure}
Figure~\ref{fig:AIA_CHbound_Compare} shows the comparison of the CH boundary lines, the CH boundary regions (defined in Section~\ref{subsec:DEM}) and the spatial irregularities obtained from three different CH boundary thresholds: 30 DN~s$^{-1}$ (left panel), 35 DN~s$^{-1}$ (middle panel), and 40 DN~s$^{-1}$ (right panel). Overall shapes of derived CH boundaries do not differ substantially across the threshold ranges of $\pm$ 5 DN~s$^{-1}$ of the optimal thresholds, especially in the leading and bottom boundaries. Although the trailing boundary shows more considerable changes in boundary shape, the main conclusion that the trailing boundary is more irregular/corrugated than the leading boundary is still valid. We also repeated the analysis in Section~\ref{subsec:TDplot} with the different CH boundaries and found that the results of different boundary thresholds also do not differ substantially and are qualitatively in agreement. Hence, we argued that our chosen CH boundary and the results obtained from it are robust even when subjected to small intensity variations.

\section{Extreme Ultraviolet Intensity of Coronal Hole Boundary Regions}\label{S-A2}
\begin{figure}[h!]
    \centering
    \includegraphics[width=\textwidth]{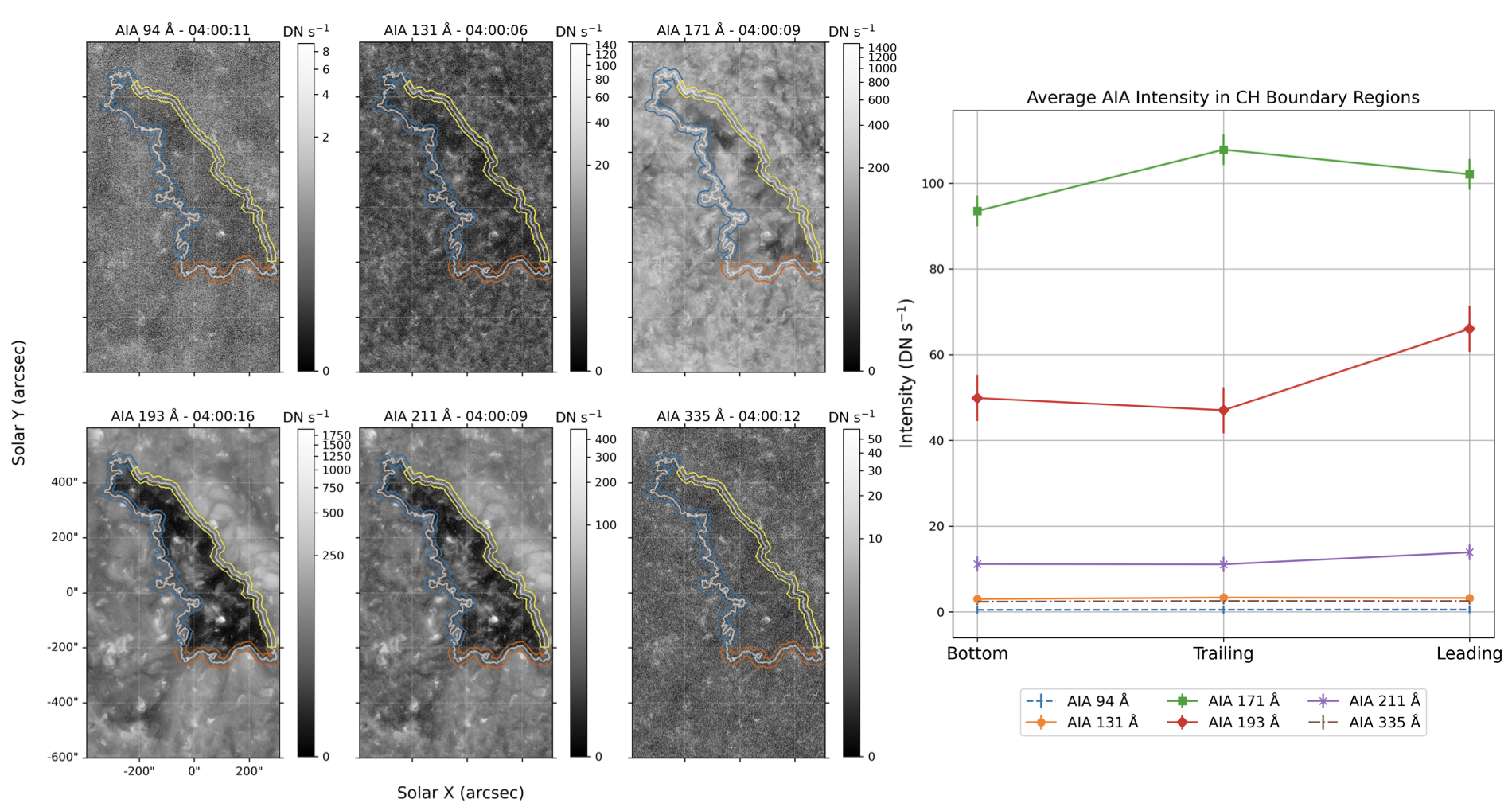}
    \caption{AIA observations of CH in six EUV passbands. The left panel shows the low-latitude CH with its boundary overplotted in white. The coloured contours indicate different sections of the CH boundary region, similar to Figure \ref{fig:DEM_1day}. The right panel shows the average EUV intensity of bottom, leading and trailing boundaries. The error bars correspond to the instrumental uncertainties.}
    \label{fig:AIA_CHBR}
\end{figure}

The left panel of Figure \ref{fig:AIA_CHBR} shows the AIA observations in 94 {\AA}, 131{\AA}, 171 {\AA}, 193 {\AA}, 211 {\AA} and 335 {\AA} passbands. Note that the data have been prepared as described in Section \ref{S-obs}. The average EUV intensities of the bottom, trailing, and leading boundaries are shown in the right panel. The distinct difference in  EUV intensities can be seen in the 171 {\AA}, 193 {\AA} and 211 {\AA} passbands. In particular, the trailing boundary has a higher average intensity in 171 {\AA} and a lower average intensity in 193 \AA\ than the leading boundary. This is consistent with the results of the DEM analysis reported in Section~\ref{subsec:DEM}.
  
%%% BIBLIOGRAPHY %%%%%%%%%%%%%%%%%%%%%%%%%%%%%%%%%%%%%%%%%%%%%%%%%%%%%%%%%%%
     % format of references provided by the journal (.bst)
\bibliographystyle{spr-mp-sola}
     % name your Bibtex file containing your references (.bib)
\bibliography{bibliography_new}  

     % Checking: look if the file containing the ``\bibitem'' exits
     %           so check if the .bbl file exist (bibTeX compilation)
\IfFileExists{\jobname.bbl}{} {\typeout{}
\typeout{****************************************************}
\typeout{****************************************************}
\typeout{** Please run "bibtex \jobname" to obtain} \typeout{**
the bibliography and then re-run LaTeX} \typeout{** twice to fix
the references !}
\typeout{****************************************************}
\typeout{****************************************************}
\typeout{}}

\end{document}